\documentclass[useAMS,usenatbib]{mn2e}

\usepackage{graphicx}
\usepackage{subfigure}

\newcommand{\hi}{H$\,${\sc i}} 
\newcommand{\ci}{C$\,${\sc i}}

\newcommand{\mjb}{mJy beam$^{-1}$}

\begin{document}

\title[Galactic HI Structure in the 
Direction of 3 distant galaxies]{Galactic HI on the 50-AU scale in the direction of three extra-galactic sources observed with MERLIN}

\author[Goss et al.]{W. M. Goss$^1$,
 A. M. S. Richards$^2$,
T. W. B.\, Muxlow$^3$,
P. Thomasson$^3$\\
$^1$National Radio Astronomy Observatory, Socorro, NM, U.S.A.  87801\\
$^2$Jodrell Bank Centre for Astrophysics, Alan Turing Building,
University of Manchester, M13 9PL, UK\\
$^3$MERLIN/VLBI National Facility, Jodrell Bank Observatory, The University of 
Manchester, Macclesfield, Cheshire, SK11~9DL, UK}

\date{Accepted []. Received []. Original form
31/12/2007}

\maketitle
\label{firstpage}

\begin{abstract}
We present MERLIN observations of Galactic 21-cm \hi\/ absorption at
an angular resolution of $\sim0.1-0.2$ arcsec and a velocity
resolution of 0.5 km s$^{-1}$, in the direction of three moderately
low latitude (--8\degr\/ $<b<$ --12\degr) extragalactic radio sources,
3C~111, 3C~123 and 3C~161, all of which are heavily reddened.  \hi\/
absorption is observed against resolved background emission sources up
to $\sim2$ arcsec in extent and we distinguish details of the opacity
distribution within 1--1.5 arcsec regions towards 3C~123 and
3C~161. This study is the second MERLIN investigation of small scale
structure in interstellar \hi\/ (earlier work probed Galactic \hi\/ in
the directions of the compact sources 3C~138 and 3C~147). The
0.1-arcsec scale is intermediate between \hi\/ absorption studies made
with other fixed element interferometers with resolution of 1 to 10
arcsec and VLBI studies with resolutions of 10--20 milli-arcsec.  At a
scale of 1 arcsec (about 500 AU), prominent changes in Galactic \hi\/
opacity in excess of 1--1.5 are determined in the direction of 3C~161
with a signal-to-noise ratio of at least 10$\sigma$. Possible
fluctuations in the \hi\/ opacity at the level of about 1 are detected
at the $2.5-3\sigma$ level in the direction of 3C~123.

\end{abstract}

\begin{keywords}{radio lines: ISM -- radio continuum: galaxies --
ISM: kinematics and dynamics -- ISM: structure -- Galaxy: kinematics
and dynamics}
\end{keywords}

\section{Introduction} 
In order to understand the dynamics of the interstellar medium (ISM),
a thorough understanding of the physical conditions of the ISM gas is
required over a large range of physical scales.  In particular, the
magnitude of changes in density of the neutral atomic component (\hi)
over a wide range of physical scales from tens of AU to parsecs
plays a major role in determining the evolution of the dense component
of the ISM. A recent conference in Socorro, New Mexico (USA) in May
2006 was devoted to this broad topic of small scale structures in
various ionized and neutral (atomic and molecular) phases of the ISM
(the SINS conference: `Small Scale Ionized and Neutral Structures';
\citealt{hg07}). A major topic of discussion at this conference was
the nature of \hi\/ structures with scales from tens to thousands of
AU. To date, most of the effort in studying the small scale structure
in \hi\/ has consisted of VLBI and pulsar determinations at the 10 to
100 AU scale (\citealt{fgdt}; \citealt{fg01}; \citealt{brog05};
\citealt{brog07}; \citealt{ws07}; \citealt{swh03}).

A major gap between the 100-AU and 0.1-pc ($2\times10^4$ AU) scales
has remained largely unexplored; \citet{ddg96} used the MERLIN array
and the European VLBI Network (EVN) to observed Galactic \hi\/
absorption towards the extragalactic sources 3C~138 (Galactic
coordinates $l$ = 187.4\degr\/ and $b$ = --11.3\degr) and 3C~147
(Galactic coordinates $l$ = 161.7\degr\/ and $b$ = 10.3\degr) over
angular scales of 50 to 1000 milli-arcsec (mas). Prominent changes in
\hi\/ opacity at the level of 0.15 were observed in the MERLIN
0.1-arcsec resolution data and at the level of 0.5 in the 50-mas EVN
data. The changes were observed over linear scales of $\la 500$ AU
arising from nearby \hi\/ gas ($<$ 500 pc).

A number of discussions of the observational characteristics of the
small scale neutral gas in addition to \hi\/ were presented at the
SINS conference \citep{l07}. \citet{aml01} have reviewed the nature of
both spatial and temporal variations in optical IS lines, while
\citet{jt07} have reviewed \ci\/ UV observations suggesting
overpressure in some portions of the ISM. If the observations of
opacity variations are interpreted in terms of density fluctuations, a
major problem of excess density and pressure is implied. To overcome
this problem \citet{heiles97} (see also \citealt{heiles07} and
\citealt{hs07}) has suggested several geometrical explanations that
would reduce the overpressure; examples include sheet or filament
morphologies.  \citet{desh00} (also see \citealt{desh07}) has, on the
other hand, proposed that the structures on small scales of tens of AU
are only a natural extension of irregularities in the distribution of
\hi\/ scales at much larger dimensions, based on a power law (`red
spectrum') derived by \citet{ddg00} using VLA data in the direction of
the Galactic source Cas~A. This derivation consisted of a power law
over scales of several parsec to less than 0.1 pc. The pros and cons
of these contrasting proposals are discussed by \citet{hs07}.

In order to fill in the gap between the 0.1-pc and 10-AU scales,
additional observations of the 100-AU scale are required. The MERLIN
array at 21 cm is the ideal instrument for this endeavour; we
undertook \hi\/ absorption observations in the direction of strong
extragalactic sources with total angular sizes of order 1
arcsec. Three sources with Galactic latitudes in the range --8\degr\/
to --12\degr\/ were observed in 2001 and 2003. These observations
represent the second instalment of Galactic \hi\/ studies following
\citet{ddg96}.

The choice of sources was guided by the pioneering (and
under-recognized) ground-breaking Galactic \hi\/ absorption studies by
\citet{ldg82} and \citet{gl86}. Observations were made in the early
1980s, using the newly completed VLA with an angular resolution of
3--4 arcsec and a velocity resolution of 1.3--1.6 km s$^{-1}$, in the
directions of 3C~111, 3C~123, 3C~161 and 3C~348. The new MERLIN
observations were made with an angular resolution of $\sim0.1-0.2$
arcsec and a velocity resolution of 0.49 km s$^{-1}$, in the direction
of all of these sources except for 3C~348.  Preliminary MERLIN results
were presented by \citet{fgm07}.

We outline the data acquisition process in Section~\ref{obs} and
describe and interpret the \hi\/ absorption towards each source in
Section~\ref{results}.  The results and implications for the structure
and kinematics of Galactic \hi\/ are summarised in Section~\ref{concl}
and we add an Appendix \ref{appendix} on hitherto unpublished data for
3C~161.

\section{Observations and Data Reduction}
\label{obs}
The observations were carried out in a two-stage process.  The
 structure of our targets had never been investigated previously at
 0.1-arcsec resolution (with the exception of the 3C123 hotspots at 5
 GHz, \citealt{hapr97}) so we started by observing the radio continuum
 from each source, at a wavelength of either 6 or 21 cm, in order to
 determine the size of the field required for the final \hi\/ line and
 continuum images. The continuum observations were made using the
 MERLIN array including the antennas at Defford, Cambridge, Knockin,
 Darnhall, Tabley, and the MK2 telescope at Jodrell Bank, giving a
 usable bandwidth of 13--15 MHz. The observing parameters are given in
 in Table~\ref{obspar}. Standard observing and data reduction
 procedures were used, see the MERLIN User Guide \citep{MUG}. The
 resulting continuum images are in good agreement with the somewhat
 lower-sensitivity images made during the \hi\/ observations (see
 below). The continuum emission from 3C161 is discussed further in
 Appendix~\ref{appendix}.

The observing parameters for the \hi\/ observations are given in
Table~\ref{obspar}.  The antennas listed above were used with the
addition of the Lovell Telescope.  Observations of the spectral
targets were interleaved with observations of nearby compact phase
reference sources listed in Table~\ref{obspar}.  We used a 0.5-MHz
bandwidth divided into 256 channels for the \hi\/ targets and the
bandpass calibration source, giving a velocity channel separation of 0.41 km
s$^{-1}$. The target data were recorded adjusted to a fixed velocity
with respect to the local standard of rest ($V_{\rm LSR}$), given in
Table~\ref{obspar}.  Other calibration sources were observed using the
standard wide-band continuum configuration for maximum sensitivity.
The absolute flux density scale for all sources was
derived from 3C~286. We used 3C~84 to calibrate the bandpass, after
eliminating its weak Galactic absorption.  We applied these solutions
and the phase and amplitude calibrations derived from the phase
reference sources to the relevant targets. More details of MERLIN
spectral line data reduction are given in \citet{MUG}.

After calibration, we inspected the spectrum of each target and identified
the line-free channels.  These were subtracted from the spectra to enable
us to make \hi\/ absorption line cubes and we also imaged the continuum.
The final
  optical depth cubes were calculated from the continuum image and the
  line cube. We used cutoffs of a few per cent of the peak continuum
  intensities (see Table~\ref{obspar} and captions to the relevant
  figures). All positions are given in J 2000 (ICRS).
We can only measure \hi\/ absorption on the same scale as
 the continuum structures, since \hi\/ emission is undetectable on
 the (sub-)arcsec MERLIN scales.  The maximum scale of emission which
 can be imaged reliably from our observations is 2--3 arcsec at 1.4
 GHz and 0.7--1 arcsec at 5 GHz. Detectable emission on slightly
 larger scales would cause a distinctive 'bowl' artefact due to
 missing flux density on most baselines.  No signs of this are
 detected in any of the MERLIN images, implying that there is minimal
 missing flux density on the scales from 0.1 to 3
 arcsec imaged here. In addition we compare the current MERLIN results with
 existing VLA data to confirm the minimal effect of missing short
 spacings. Detailed comments are presented for each source.

\section{Results}
\label{results}
\subsection{3C~111}

\begin{figure*} \centering
  \includegraphics[angle=0,width=18cm]{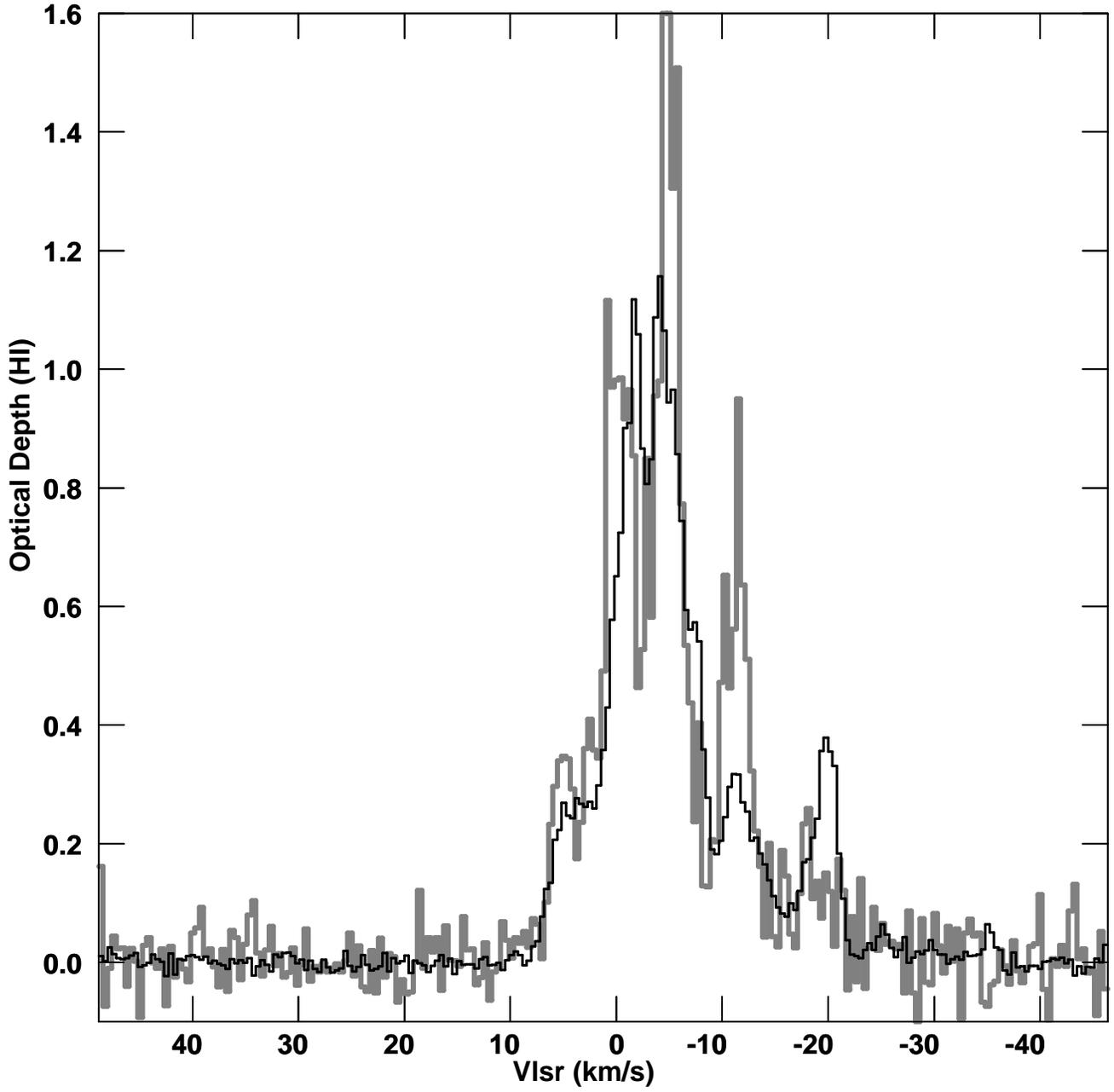} 
\caption{The \hi\/ optical depth
spectra towards the core and the resolved NE component of 3C111 are shown
by the thin black line and the thick grey line, respectively.  The
velocity channel separation is 0.41 km s$^{-1}$ and velocities are given
with respect to the Local Standard of Rest. The
optical depth was averaged over areas of 0.3 arcsec $\times$ 0.3
arcsec for the core and 0.45 arcsec $\times$ 0.65 arcsec for the NE component.
\label{3c111tauspec}} \end{figure*}

The large low latitude source 3C~111 ($l$ = 161\fdg68, $b$ =
--8\fdg82) has a complex continuum structure. This FR (Fanaroff-Riley)
II source is a nearby broad line radio galaxy at a redshift of
0.049. VLA continuum studies were carried out by \citet{lp84};
\citet{gl86} have also summarized VLA \hi\/ observations of this
source, identifying a core continuum component (their `2' ) and two
prominent lobes with hotspots, NE (their `1', displacement 2~arcmin)
and SW (their `3' and `4', displaced by 80~arcsec).  
Using the MERLIN
array, the NE component is detected with an extent of $\sim2.5$ arcsec
and a continuum peak of only 46 mJy~beam$^{-1}$ (not listed in
Table~\ref{obspar}); the mean \hi\/ spectrum was obtained after
smoothing to a 0.3-arcsec resolution. 
The core of the galaxy has a
peak of 1.27 Jy~beam$^{-1}$. The two \hi\/ spectra for components 1
and 2 are superimposed in Fig.~\ref{3c111tauspec}.

The SW component is
heavily resolved by MERLIN; the emission detected over $\sim2$ arcsec
has a peak of only 8 mJy beam$^{-1}$. At this level, detection of
\hi\/ absorption is not possible even after some smoothing. We also
note that the highest apparent \hi\/ opacities for the NE component
(at --5 and 0 km s$^{-1}$) are unreliable with uncertainties in excess
of 1, due to the partial saturation of the \hi\/ line and the weakness
of the resolved continuum feature. 
The faintness of the lobes of 3C111 implies that it is
impossible to determine detailed \hi\/ opacity images.  The spectra
obtained for the core source and for the average of the NE lobe using
MERLIN are in excellent agreement with earlier VLA data obtained with a
resolution of 3.5 arc sec obtained by
\citet{gl86}.
We confirm the increase in the \hi\/
opacity for the --19 km s$^{-1}$ feature, from $\tau$ \hi\/ $\sim$
0.35 towards the core (component 2), compared with $\tau$ \hi\/ $<$
0.1 towards the NE hotspot (component 1). A relatively higher opacity
of $\tau$ \hi\/ $\approx0.6$ is observed towards the NE hotspot at
--11 km s$^{-1}$, compared with $\tau$ \hi\/ $\approx0.3$ towards the
core.  

\citet{mm95} made observations towards 3C~111 using the VLA at
16--14-arcsec resolution at the respective \hi\/ and OH line
frequencies of 1.4 and 1.6 GHz as well as multi-epoch observations of
H$_2$CO at 4.8 GHz in various configurations. Their images are also
consistent with the \hi\/ results from \citet{gl86}, containing
substantial absorption line variations in \hi\/ and OH which suggest
with structure on scales of 0.14 to 0.34 pc at their adopted distance
of 350 pc.  The time variability in H$_2$CO absorption observed
towards the core by \citet{mm95} and \citet{mmb93} over a 3.4 yr
period provides additional evidence for 10-AU concentrations with
number densities $>10^6$ cm$^{-3}$.  These apparent fluctuations could
arise from
a combination of the elliptical apparent motion due to parallax,
linear solar motion and the proper motion of the cloud \citep{mm95},
giving a transverse drift of about 4 AU per year (at the adopted
distance of 350 pc) of the position of the line of sight through each
cloud towards the background source.

\subsection{3C~123} 
\label{sec:3C123}
\begin{figure*} \centering \includegraphics[angle=0,width=18cm]{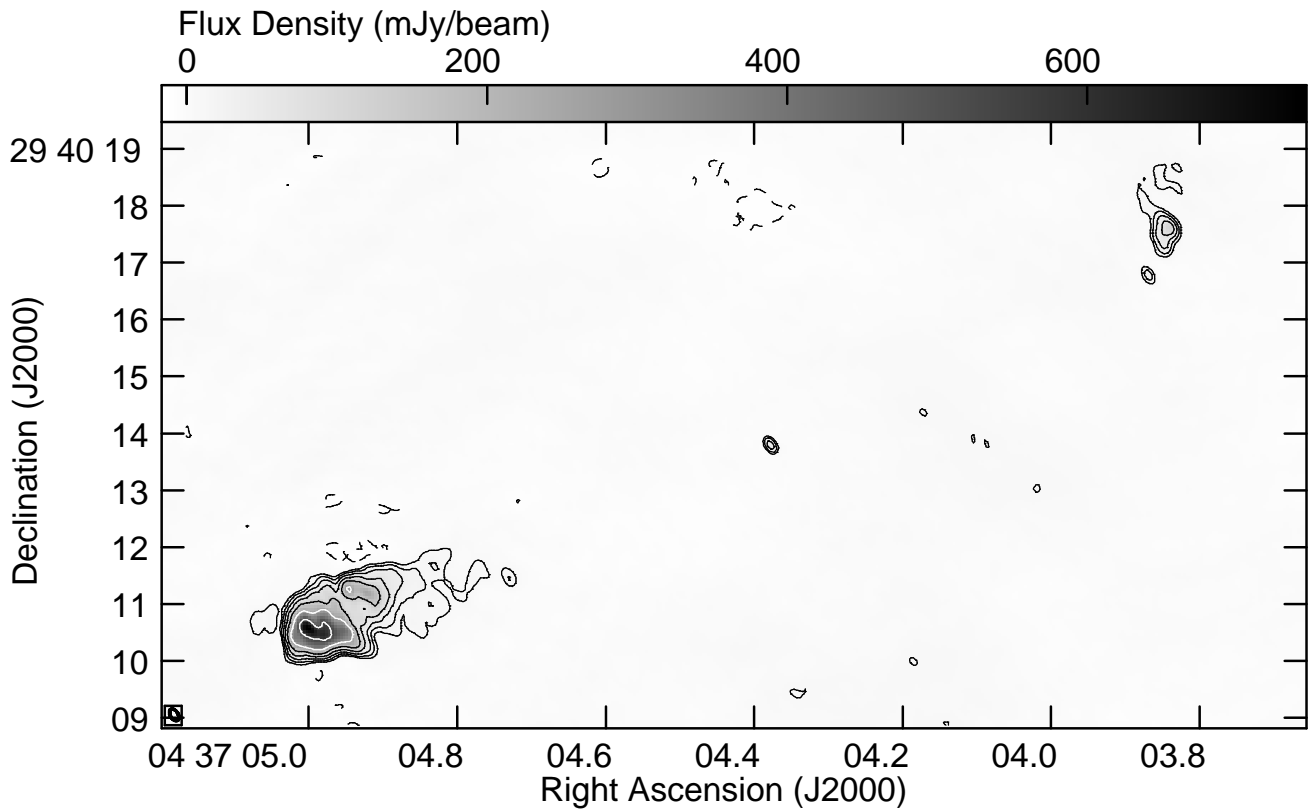} \caption{MERLIN continuum image of
3C~123 observed at 1408 MHz, using 13.5-MHz bandwidth. The beam is
0.23 arcsec $\times$ 0.14 arcsec at a position angle of 28\degr,
indicated in the lower right corner.  
The contour levels are (-3, 3, 6, 12, 24, 48, 96, 192) $\times$ the
rms noise level of 2.6 mJy beam$^{-1}$.  The weak core is detected as
well as the two prominent hotspots.  The greyscale shows the flux density in mJy beam$^{-1}$.
 \label{3c123cont}} \end{figure*}

The continuum image of 3C~123 ($l$ = 170\fdg58, $b$ = --11\fdg66) is
shown in Fig.~\ref{3c123cont}. Previous high-resolution VLA images
of this FR II radio galaxy have been presented by \citet{cgs91}
(images made by R. Laing at 1.7, 4.9 and 14.9 GHz) and by
\citet{hapr97}. 3C~123 has been identified with a 19 mag galaxy with
strong Galactic reddening at a redshift of 0.22. \citet{hapr97} show a
5 GHz MERLIN image (beam 0.05~arcsec); both components shown in
Fig.~\ref{3c123cont} (the E and W hotspots) have been imaged by these
authors. The weak core ($\sim 50$ mJy) is also detected in the current
image at right ascension $04^{\rm h} 37^{\rm m} 04\fs38$, declination
$29\degr 40\arcmin 13\farcs8$ (J2000).
The MERLIN continuum image of 3C123 contains all the flux
density that was detected by the VLA at 1.4 GHz with 1.2 arcsec
resolution (R. Laing in \emph{Atlas of DRAGNS}
http://www.jb.man.ac.uk/atlas/). The total flux density of 13.5 Jy
above $3\sigma$ in the MERLIN image of the E hotspot
(Fig.~\ref{3c123cont}) in an area of about 10 arcsec$^2$ is in good
agreement with the VLA image over the same region. The central core
radio source is unresolved by MERLIN and has a flux density of 48 mJy,
compared to 51 mJy in the VLA image.

\begin{figure*} \centering \includegraphics[angle=0,width=18cm]{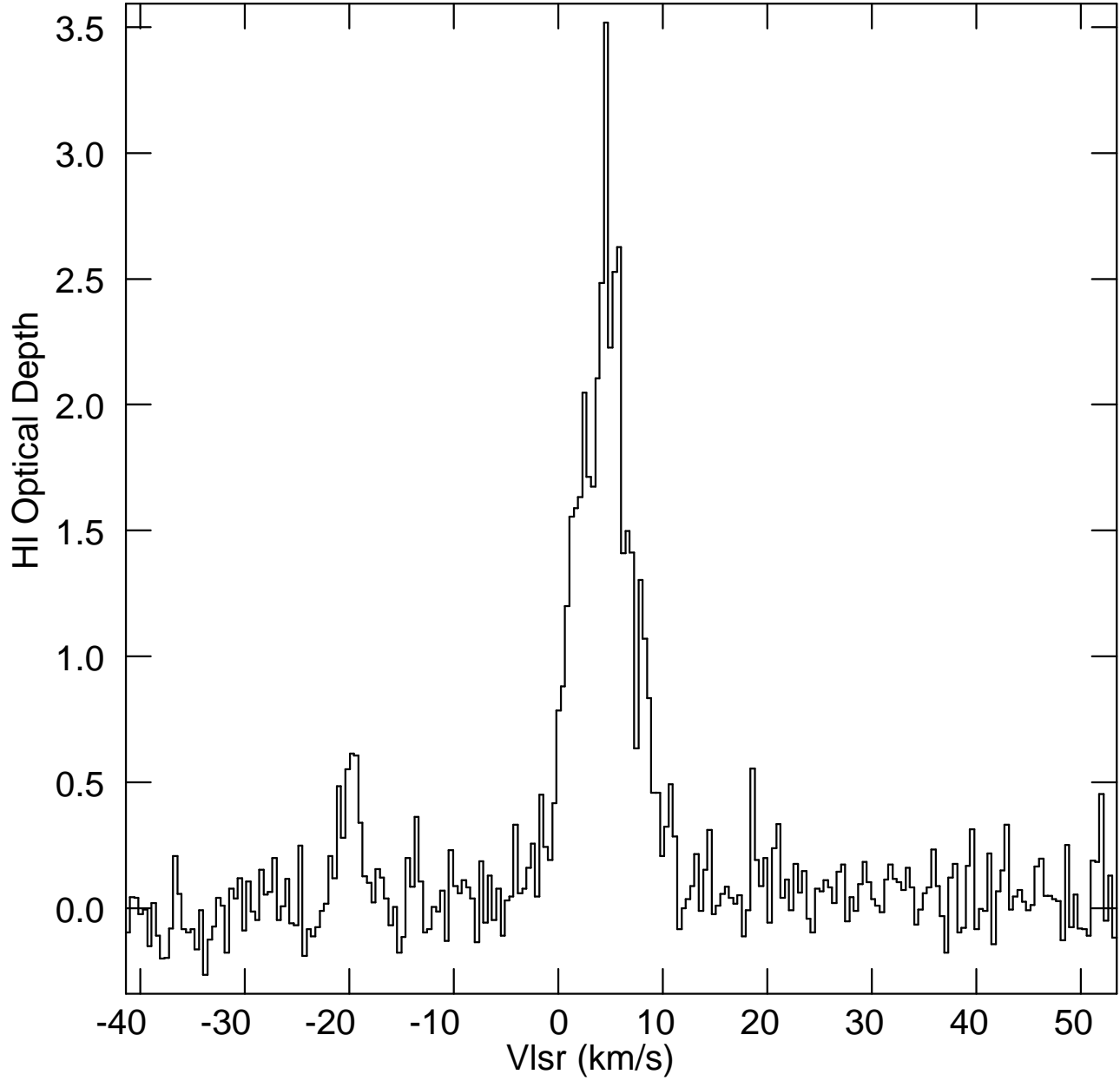} \caption{Mean \hi\/ optical depth
spectrum towards the peak emission component (SE) in 3C~123 as observed with 
MERLIN, averaged over an area of $0.25$ arcsec $\times$ $0.25$ arcsec
centred at 04$^{\rm h}$ 37$^{\rm m}$ 05\fs01  +29\degr 40\arcmin
11\farcs0 (J~2000).   The
velocity channel separation is 0.41 km s$^{-1}$ and velocities are given
with respect to the Local Standard of Rest.
\label{3c123tauspec}} \end{figure*}

The mean \hi\/ optical depth spectrum toward the SE
hotspot (over an area of 0\farcs25 $\times$ 0\farcs25) is shown in
Figure~\ref{3c123tauspec}. This spectrum is similar to the results of
\citet{ldg82}. Their anomalous velocity feature at --73 km s$^{-1}$ is
outside the velocity range ($\pm \approx$50 km s$^{-1}$) of the
current observations. The narrow line at --20.1 km s$^{-1}$ (FWHM
$2.9\pm0.2$ km s$^{-1}$) and the +4.2 km s$^{-1}$ feature (FWHM
$6.0\pm0.5$ km s$^{-1}$) are prominent; the latter has been associated
with a CO line indicative of a possible association with a molecular
cloud (\citealt{ldg82}; \citealt{lb79}). The mean opacity of the
--20.1 km s$^{-1}$ line in the MERLIN data is somewhat larger than the
value reported by \citet{ldg82} (their position `c'); the properties
of the +4.2 km s$^{-1}$ feature agree closely.  In the absence of a
direct measurement, a possible distance $d$ is given by
\begin{equation}
d= <\vert{z}\vert>/\sin{b},
\label{eq:z} 
\end{equation}
where $z$ = 107 pc is the mean value of the Galactic height
\citep{c78}. This suggests that the moderate-velocity \hi\/ observed
towards 3C123 is at $d\approx 530$ pc (as adopted by \citealt{ldg82}),
where 0.1~arcsec corresponds to transverse sizes $\approx$ 50 AU.  The
absorption at --20.1 km s$^{-1}$ probably arises from more distant gas but
a more precise estimate is not possible in this direction.

\begin{figure*} \centering \includegraphics[angle=0,width=18cm]{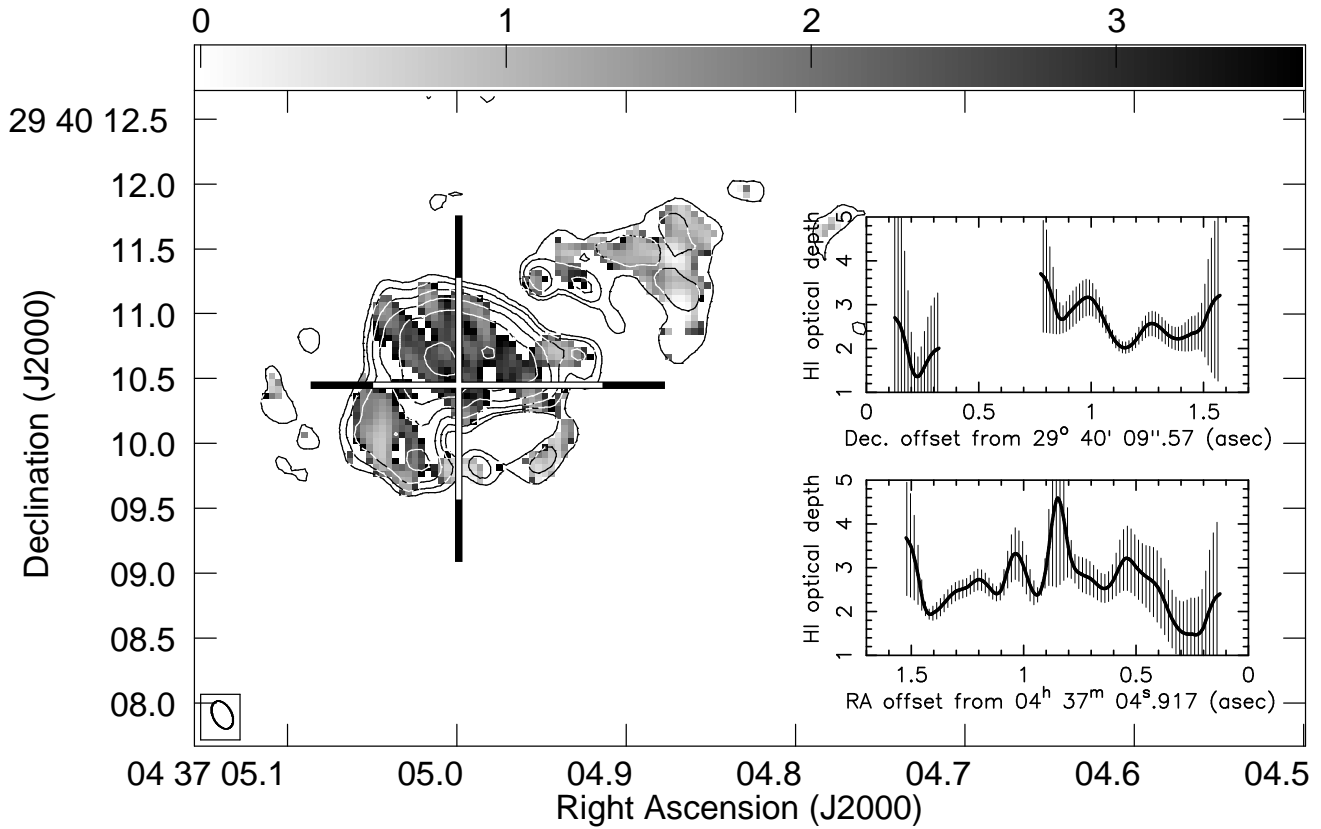} 
\caption{MERLIN \hi\/ optical depth image towards the SE component of
3C~123 in a single velocity channel centred at 4.2 km s$^{-1}$,
adopted distance to \hi\/ 530 pc.  The greyscale represents \hi\/
optical depth, and the contours represent continuum emission at (--3,
3, 6, 12, 24, 48, 96) $\times$ the rms noise level of 6 \mjb\/ for
0.15-MHz bandwidth of line-free channels at 1420 MHz.  The \hi\/
optical depth was not calculated where the continuum is $<3$ per cent
of the peak or where the \hi\/ line is completely saturated. The
inserts show slices in \hi\/ optical depth along the white portions of
the horizontal and vertical lines, from E-W (West at right, lower
plot) and N-S (south at left, upper plot) respectively. Symmetric
$\pm1\sigma$ error bars are indicated; note that at the highest
opacities the upper uncertainties may be greater than the lower.
\label{3c123taumap4}} \end{figure*}

\begin{figure*} \centering \includegraphics[angle=0,width=18cm]{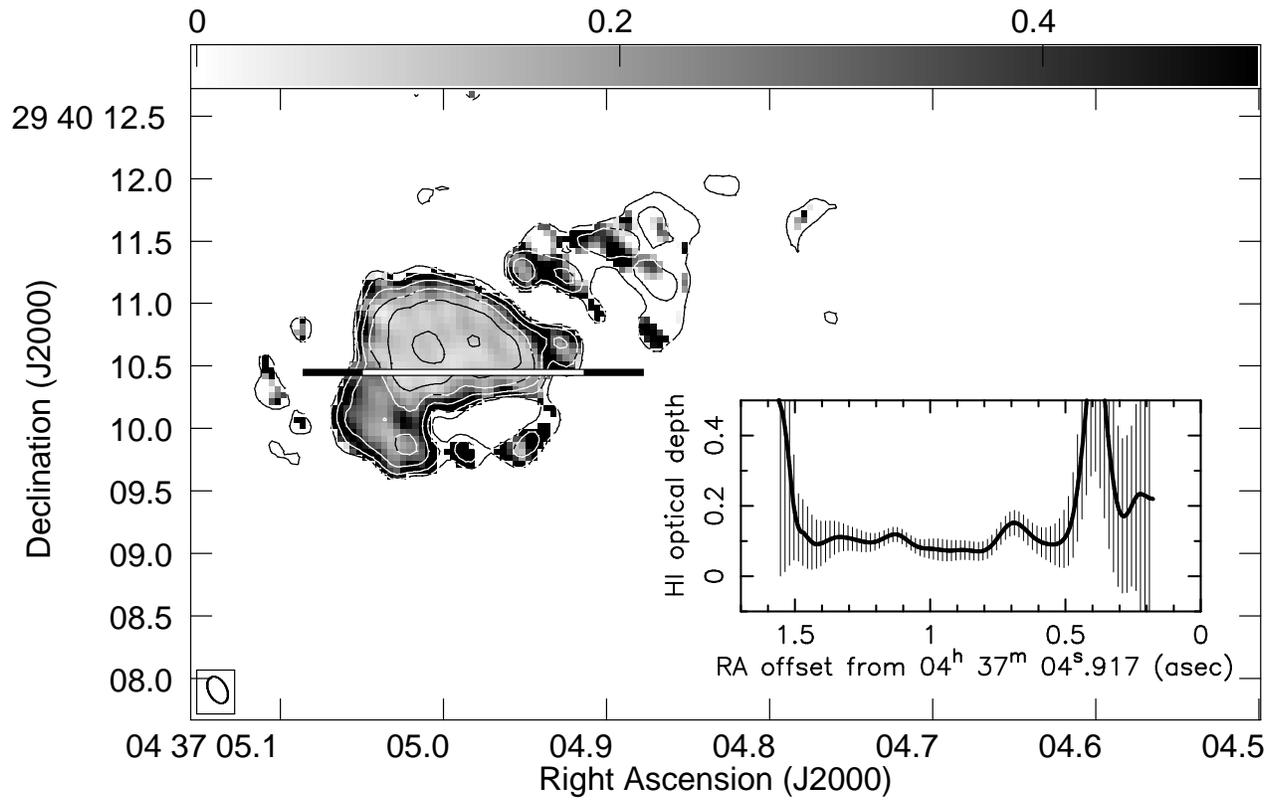} \caption{\hi\/ optical depth
image towards 3C~123 in a single channel centred at --20.1 km s$^{-1}$, adopted distance to \hi\/ $>$530 pc.
The insert shows an E-W slice. See \protect{Fig.~\ref{3c123taumap4}}
for more details.
\label{3c123taumap21}} \end{figure*}

In Figs.~\ref{3c123taumap4} and~\ref{3c123taumap21} we show the major
new results based on the MERLIN \hi\/ data for
3C~123. 
Figure~\ref{3c123taumap4} shows the optical depth image for the
strongest line at 4.2 km s$^{-1}$. At this velocity, variations in the
Galactic \hi\/ opacity are observed with a statistical significance of
only $2-3\sigma$. The signal-to-noise ratio is limited due to the low
surface brightness of this resolved source, using a beam size of
$\sim0.14$ arcsec. The angular scales of the variations vary from 0.2
to 0.6 arcsec (100 to 300 AU). In the N-S slice a variation of
$\approx1\pm0.4$ in opacity is observed.  In the E-W slice a variation
in opacity of $0.7\pm0.25$ is observed on a scale of 0.2 arcsec and a
variation of $1.5\pm0.5$ over 0.3 arcsec.
 Figure~\ref{3c123taumap21} shows a comparable plot of the variation
of \hi\/ opacity at --20.1 km s$^{-1}$. The weak absorption is
essentially constant over the scale of $\sim 2$~arcsec (1000 AU) at a
level of $\tau$(\hi) $\sim 0.1$.

\subsection{3C~161}

This low-latitude heavily reddened field ($l$ = 215\fdg44, $b$ =
--8\fdg07) remains an empty field with no certain optical
identification. 3C~161 is compact (\citealt{ujpf81}; \citealt{p82};
\citealt{ppr85}) with observed radio structure on the arcsec
scale. The later authors present a VLA image at 5~GHz (beam
0.4~arcsec) showing an obvious extension to
the West. \citet{et04} have pointed out that 3C~161 is a CSS
(compact steep spectrum source) with a spectral index of $\sim
-1.2$. The source has also been discussed by \citet{km96}. The
continuum image made during the VLA \hi\/ observations of \citet{gl86}
(beam 4.2~arcsec) remains a remarkable contribution in providing the
overall morphology of the fainter components of the source over a
range of $\sim$ 1~arcmin (see Appendix~\ref{appendix}). The source `B'
20~arcsec to the NE \citep{ujpf81} is not detected by
\citet{gl86}. However, the strongest component of 3C~161 has not been
detected by VLBI techniques; \citet{ffp00} did not detect 3C~161
during the VSOP prelaunch VLBA 5 GHz survey.  Faison and Goss have
also observed 3C~161 with the VLBA in 1998 (unpublished) with a faint
detection of emission on an angular scale of tens of mas. Thus 3C~161
contains no prominent milli-arcsec scale structure and is an ideal
candidate for MERLIN 0.1~arcsec \hi\/ imaging. 3C~161 is the brightest
of our three sources (the continuum flux density exceeds $0.1$ Jy
beam$^{-1}$ over almost 1 arcsec$^{2}$), giving very good sensitivity
to \hi\/ absorption on scales of 100 mas.

\begin{figure*} \centering \includegraphics[angle=0,width=18cm]{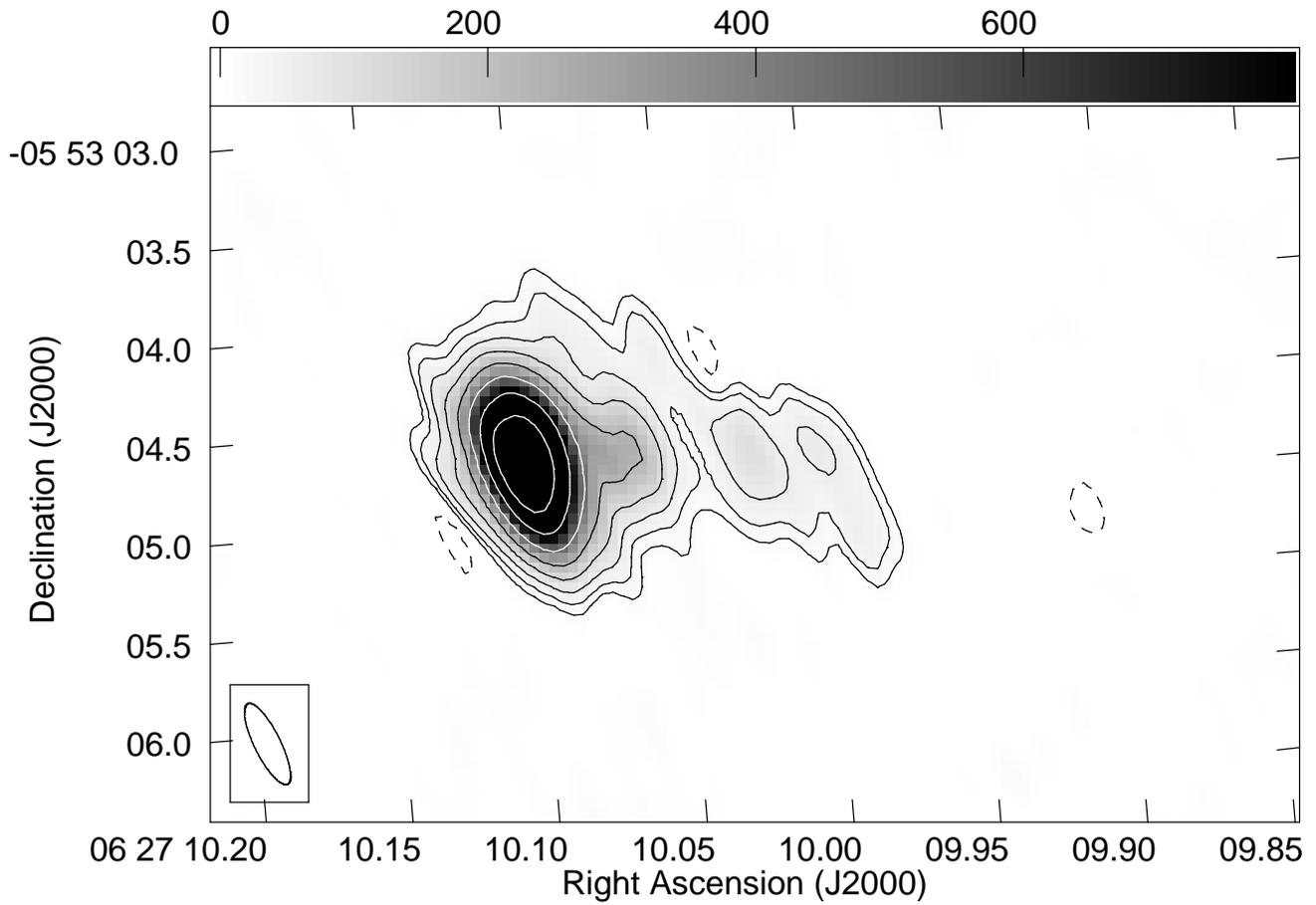} \caption{MERLIN continuum image
of 3C~161 observed at 1420 MHz in 0.18-MHz bandwidth of line-free
channels.  The image has been rotated by 4\fdg9 counter-clockwise.
The contour levels are (--3, 3, 6, 12, 24, 48, 96, 192, 384) $\times$
the rms noise level of 4 mJy beam$^{-1}$.  The beam is 0.46 arcsec
$\times$ 0.13 arcsec at a position angle of 21\degr, indicated in the lower right
corner. The greyscale shows the flux density in mJy beam$^{-1}$.
\label{3c161cont}} \end{figure*}

The continuum image obtained from the \hi\/ data in March 2002 is
shown in Fig.~\ref{3c161cont}. This and the following images have
been rotated by 4\fdg9 counter-clockwise in order to optimise the
information in optical depth slices. 

\begin{figure*} \centering
  \includegraphics[angle=0,width=18cm]{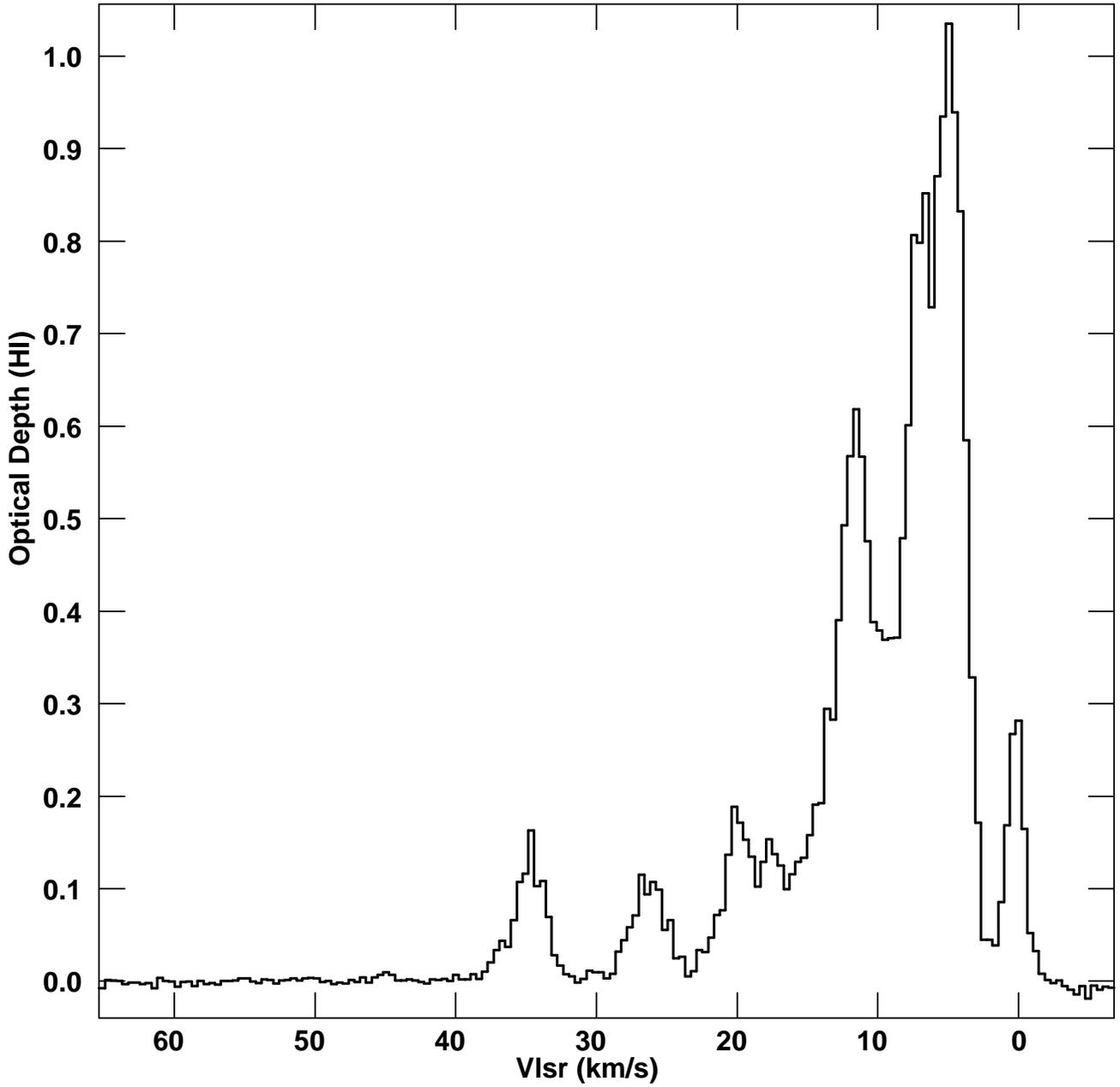} 
\caption{The \hi\/ optical depth
spectrum  towards the continuum peak of  3C~161, 
averaged over 0.35 arcsec $\times$ 0.35 arcsec.  The
velocity channel separation is 0.41 km s$^{-1}$ and velocities are given
with respect to the Local Standard of Rest. Note the wide range of velocities over which the \hi\/ absorption is observed. 
\label{3c161tauspec}} \end{figure*}

We retrieved the data presented by \citet{gl86} from the VLA archive
and reduced them in order to check whether our results were
consistent.  The total continuum flux density detected by MERLIN in
the 10-arcsec$^2$ region around component `1', shown in
Fig.~\ref{3c161cont}, is 10.5 Jy, compared to a peak flux density of
12.1 Jy in the newly constructed VLA image at (5.7 $\times$ 4.2)
arcsec$^2$ resolution.  Thus, the MERLIN continuum image contains
almost 90 per cent of the flux density as observed at lower resolution
over the relevant field of view.
  The expectation that the MERLIN line data represents most of the VLA
emission/absorption in this region is verified by the comparison of
the two integrated \hi\/ spectra over the extent of the continuum
source shown in Fig.~\ref{3c161cont}. The VLA and MERLIN data agree
remarkably well, given the uncertainties. In addition, the MERLIN
spectrum over a much smaller region of (0.35 $\times$ 0.35)
arcsec$^2$, shown in Fig~\ref{3c161tauspec}, also agrees with the mean
VLA spectrum as expected. There remains the possibility that for
3C~161 the opacities at $\tau>1.3$ may be somewhat overestimated (by
$\delta\tau \approx 0.2 - 0.3$), if there is a slight amount of missing
continuum flux density.  The weak extended components of 3C~161 (see
Fig~\ref{3C161_VLA_L.PS}) are not detected in the MERLIN data due to
the larger angular sizes and resulting lower surface brightnesses
(giving $<0.1$ mJy beam$^{-1}$ on the scale of the MERLIN beam)
compared with the bright NE lobe. There is some evidence in
Fig.~\ref{3c161tauspec} for a separate line at $\approx7$ km s$^{-1}$, close to a
line detected in the earlier NRAO 3-element interferometer data of
\citet{g76}

The observed range of velocities from 0 to 35 km s$^{-1}$
suggests the influence of Galactic rotation in the third quadrant of
the Galaxy with more positive velocities arising from IS gas at larger
distances from the Sun.  As an example the velocity of 34.2 km
s$^{-1}$ would have an inferred distance of $\sim3.5$ kpc. The
distances that can be associated with gas at lower velocities ( less
than $\sim12$ km s$^{-1}$, twice the cloud-to-cloud velocity
dispersion) are quite uncertain and a better estimate may well be to
use Equation~\ref{eq:z} with a resultant distance of 0.8
kpc. (\citet{gl86} have discussed the problem of distances of \hi\/
features in this direction. They point out that for 3C161 : `` ... the
Gould Belt phenomenon carries the cloud velocities to rotationally
permitted values..'' )  

\begin{figure*} \centering \includegraphics[angle=0,width=18cm]{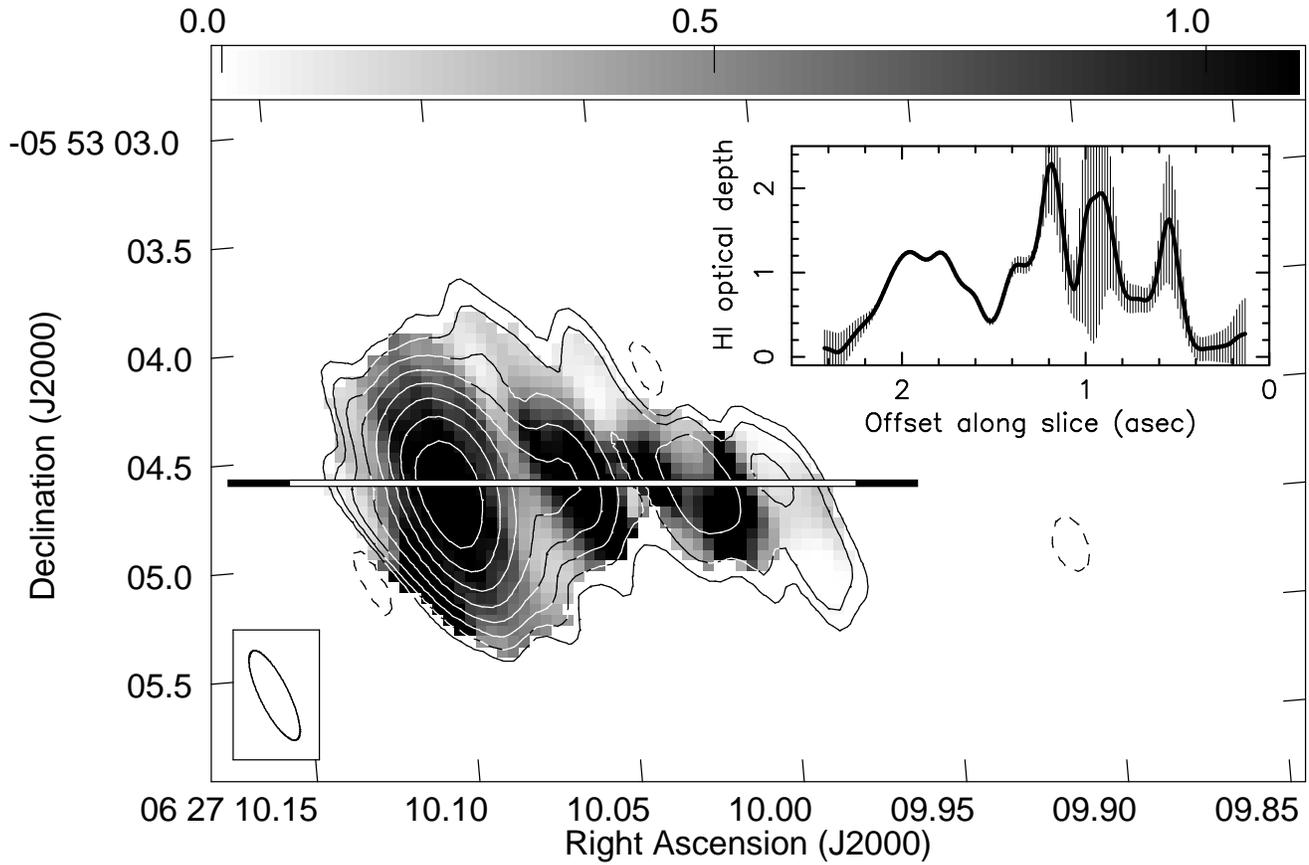} \caption{MERLIN \hi\/ optical depth
image toward 3C~161 in a single channel centred at 4.5 km s$^{-1}$,  adopted distance to \hi\/ 0.8 kpc. The greyscale represents \hi\/ optical
depth, and the contours represent continuum emission as in Fig.~\protect{\ref{3c161cont}}. 
 The \hi\/ optical depth was not calculated where the
continuum is $<1$ per cent of the peak. The image has been rotated by 4\fdg9
counterclockwise. The insert shows a slice in
 \hi\/ optical depth
along the white portion of the horizontal line (West at right).
 Symmetric $\pm1\sigma$ error bars are indicated; in some cases these
are less than the width of the line. Note that at the highest
opacities ($\tau > 1.5$), the upper uncertainty is likely to be 1.5 to
2 times larger than the lower uncertainty (which is plotted).
 \label{3c161taumap148}} \end{figure*}

\begin{figure*} \centering \includegraphics[angle=0,width=18cm]{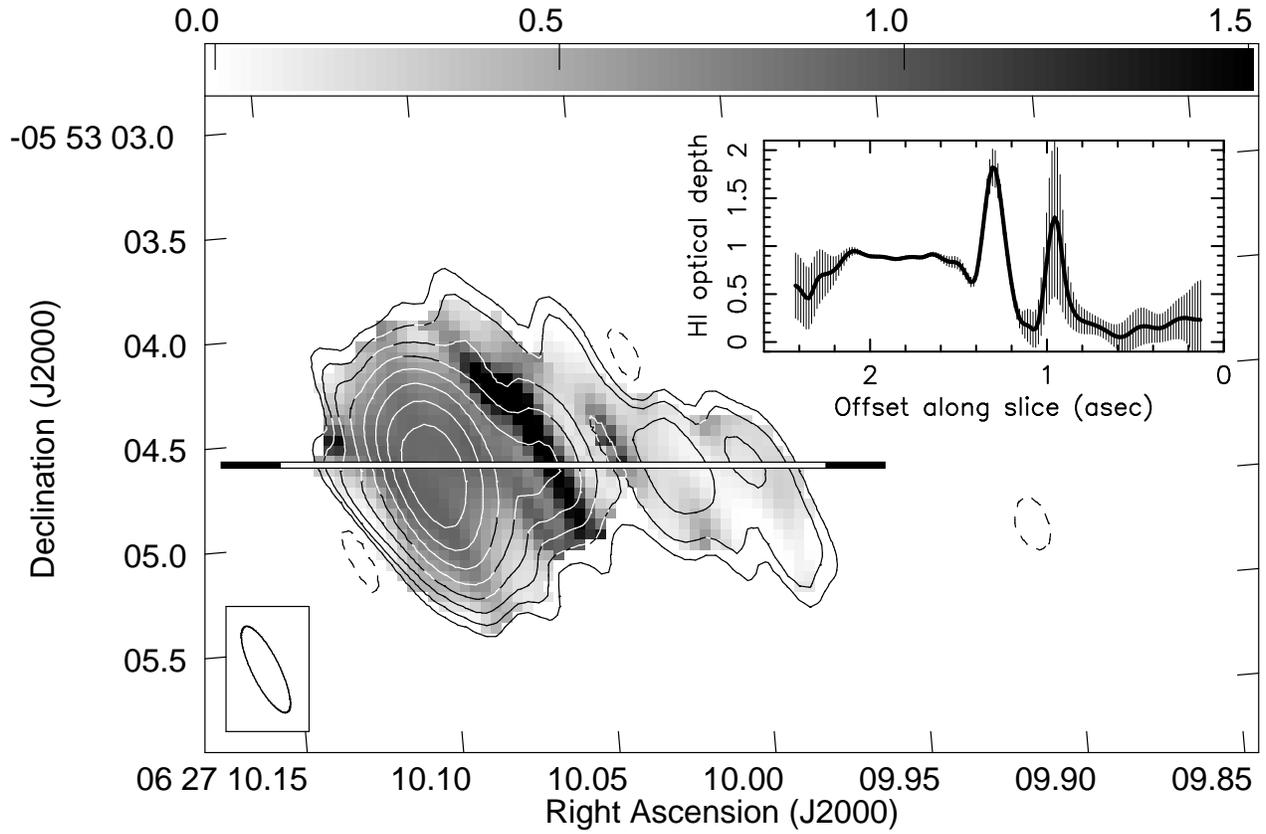} \caption{MERLIN \hi\/ optical depth
image toward 3C~161 in a single channel centred at 6.6 km s$^{-1}$,  adopted distance to \hi\/ 0.8 kpc. See Fig.~\protect{\ref{3c161taumap148}} for more details.  The opacities shown in the insert for the compact component at an
offset of 1.3 arcsec may be overestimated (by 0.2 -- 0.3), if there is a slight amount of
missing flux density in the continuum.
 \label{3c161taumap143}} \end{figure*}

\begin{figure*} \centering \includegraphics[angle=0,width=18cm]{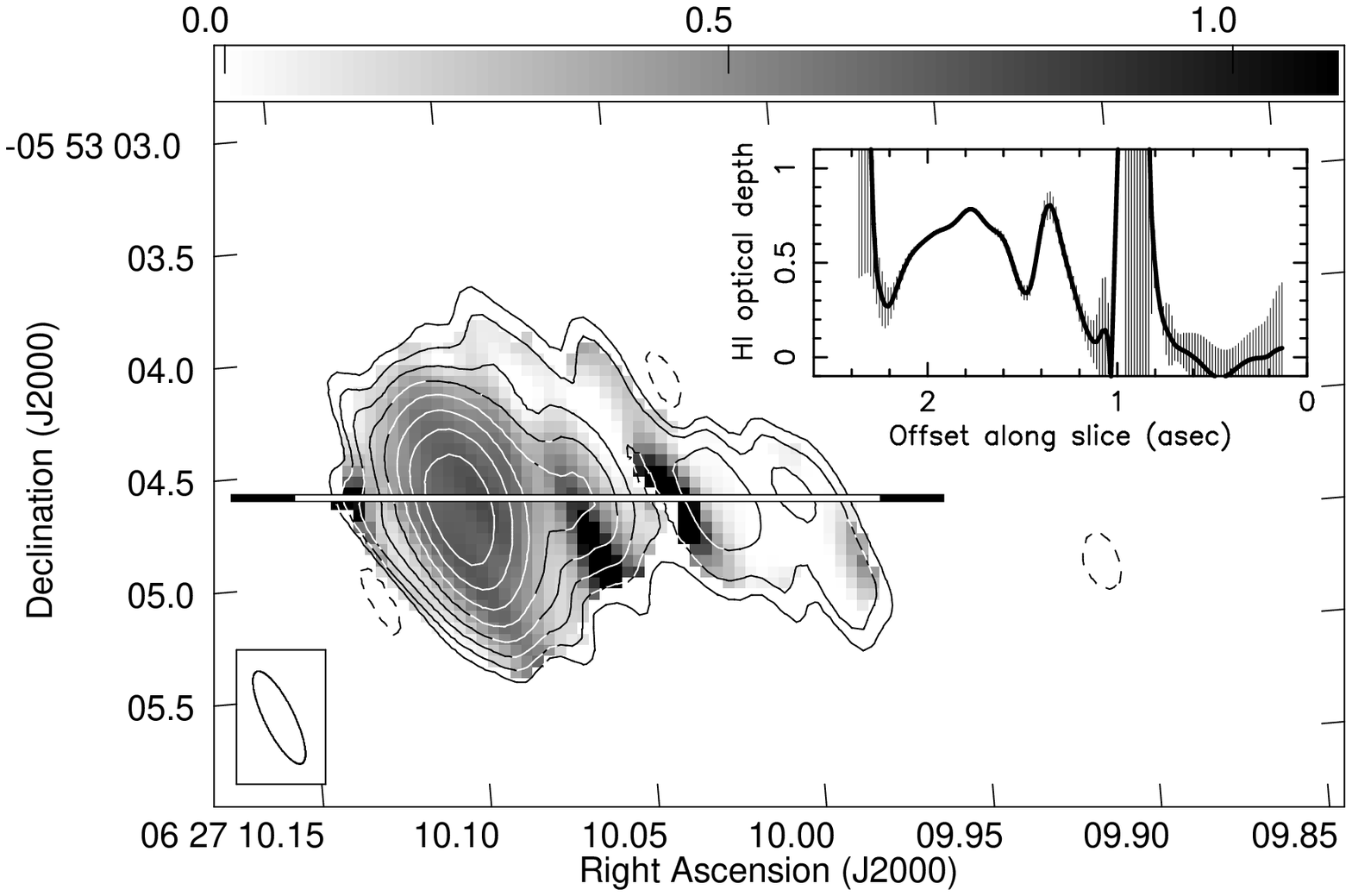} \caption{MERLIN \hi\/ optical depth
image toward 3C~161 in a single channel centred at 11.5 km s$^{-1}$,  adopted distance to \hi\/ 0.8 kpc. See Fig.~\protect{\ref{3c161taumap148}} for more details.  
 \label{3c161taumap131}} \end{figure*}

\begin{figure*} \centering \includegraphics[angle=0,width=18cm]{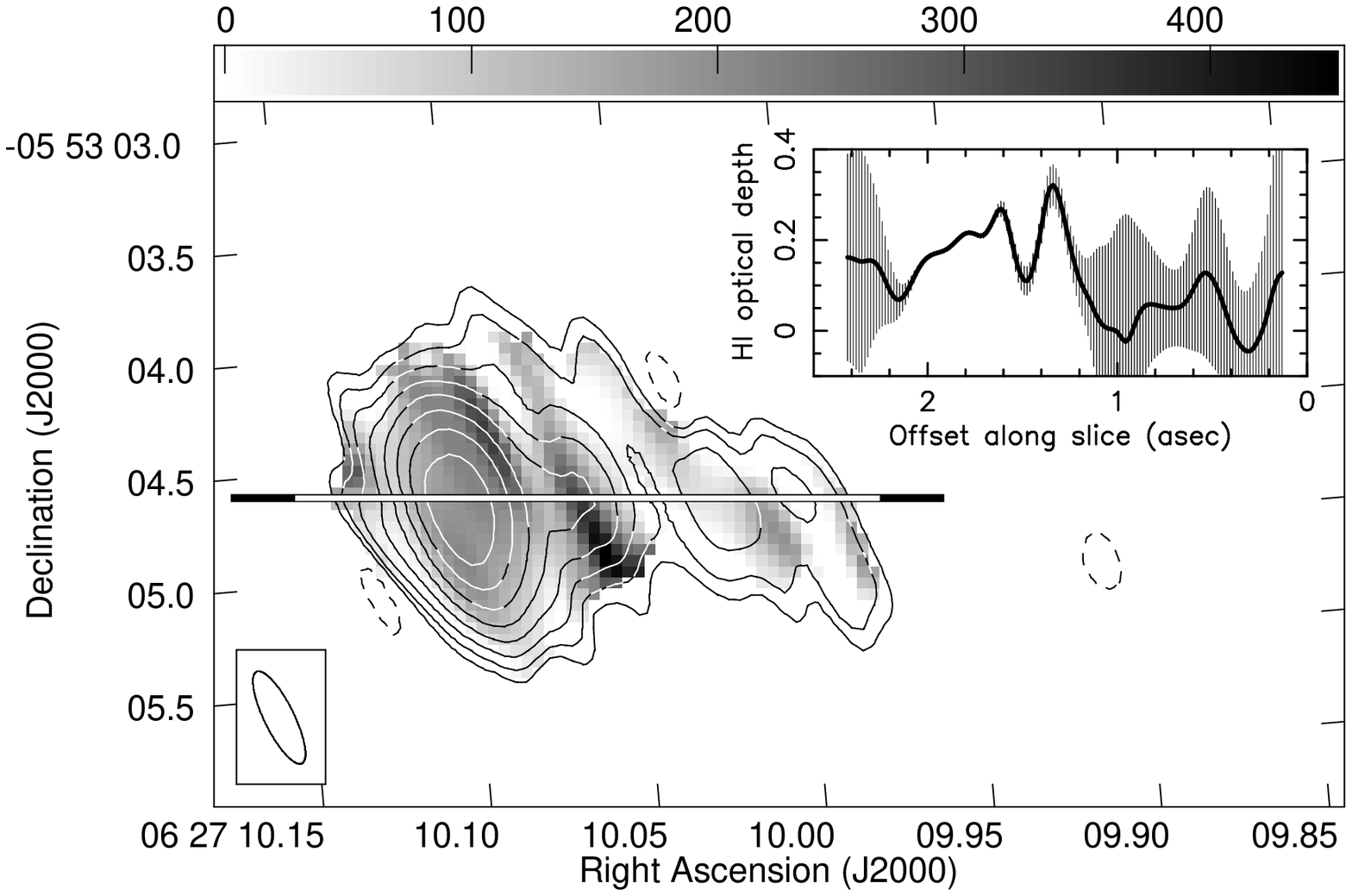} \caption{MERLIN \hi\/ optical depth
image toward 3C~161 in a single channel centred at 20.2 km s$^{-1}$,  adopted distance to \hi\/ 1.8 kpc. See Fig.~\protect{\ref{3c161taumap148}} for more details.  
 \label{3c161taumap110}} \end{figure*}

\begin{figure*} \centering \includegraphics[angle=0,width=18cm]{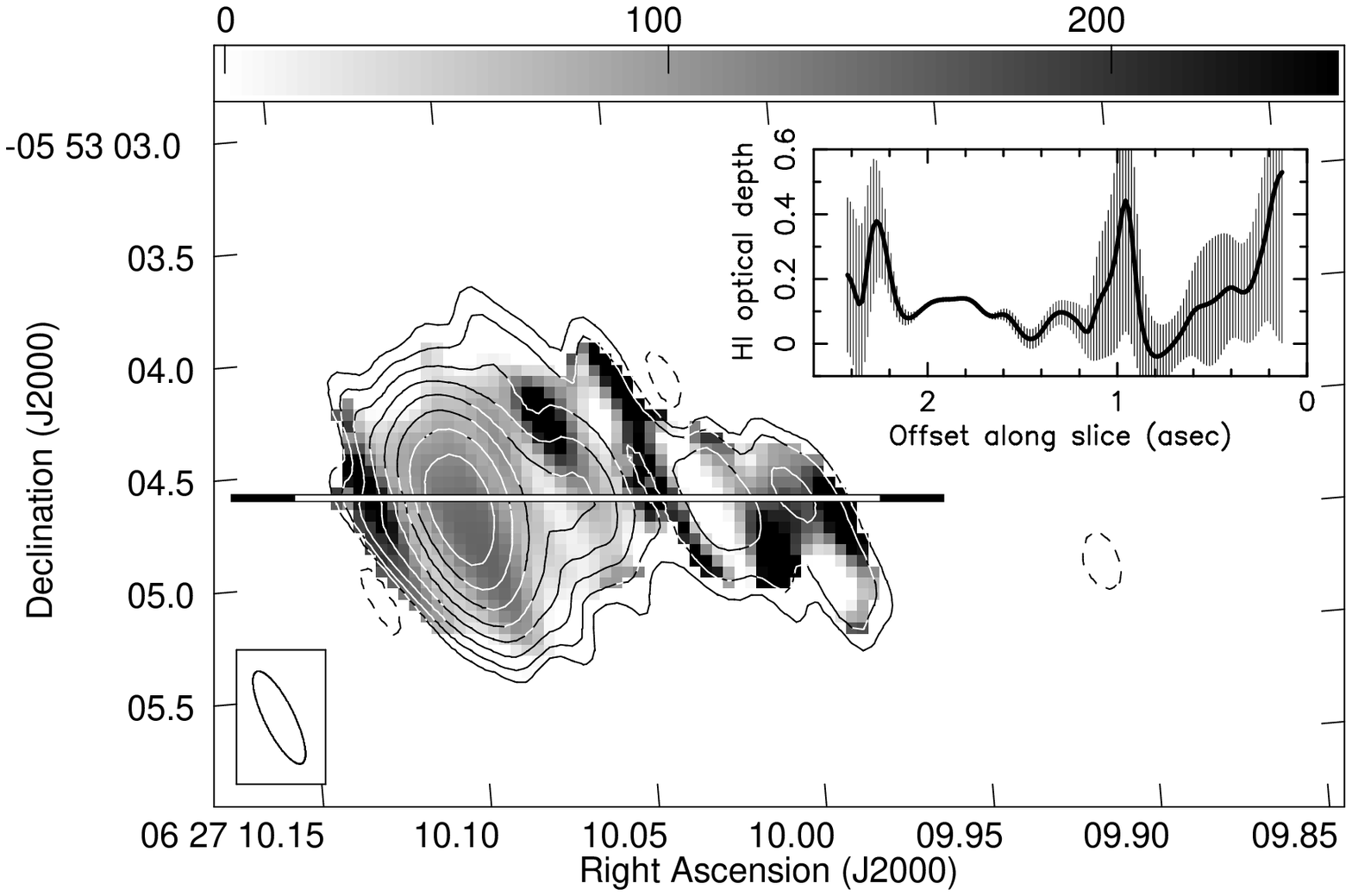} \caption{MERLIN \hi\/ optical depth
image toward 3C~161 in a single channel centred at 34.2 km s$^{-1}$,  adopted distance to \hi\/ 3.5 kpc. See Fig.~\protect{\ref{3c161taumap148}} for more details.  
 \label{3c161taumap76}} \end{figure*}

The MERLIN beam has an NS:EW axial ratio of $\sim 3.5:1$
and the resolved emission from 3C~161 is effectively only one
beamwidth wide in the N-S direction. We therefore show slices in the
direction of the long axis of the source, in the direction of
the most favourable resolution.
Figures~\ref{3c161taumap148}, ~\ref{3c161taumap143},
~\ref{3c161taumap131}, ~\ref{3c161taumap110} and ~\ref{3c161taumap76}
show optical depth images with inserts showing slices of \hi\/ optical
depth variation in single channels at velocities of 4.5, 6.6, 11.5,
20.2 and 34.2 km s$^{-1}$.  In the analysis following, we will derive
distances for the lower velocity \hi\/ features
(Figs.~\ref{3c161taumap148}, ~\ref{3c161taumap143}
and~\ref{3c161taumap131}) by assuming that they arise from gas at 0.8
kpc, where the MERLIN beam of 0.13 arcsec corresponds to about 100
AU. We will use the flat Galactic rotation curve for the higher
velocities. These estimates are all quite uncertain.

The variation of $\tau$(\hi) at 4.5 km s$^{-1}$
(Fig.~\ref{3c161taumap148}) is complex.  The strongest \hi\/
absorption, at a positional offset along the slice of 1.8 arcsec,
shows an excess opacity (with respect to the local background opacity)
at a level of $\approx1$.  This feature is spatially resolved with an angular
size of $\sim$ 0.5~arcsec, or 400 AU.  Its signal-to-noise ratio is $>30\sigma$.  The other features in the slice show opacity
fluctuations with poor signal to noise which are probably not
significant.  At a velocity of 6.6 km s$^{-1}$
(Fig.~\ref{3c161taumap143}), there is a prominent \hi\/ peak at a
displacement of 1.3 arc sec; $\tau$(\hi) fluctuates by $\approx1.4$
over a region of angular size $<0.15$ arcsec (120 AU). These two spectral
features overlap but can be decomposed into  Gaussian line
profiles, each of FWHM $2.3\pm0.2$ km s$^{-1}$.

  Farther East,
the \hi\/ distribution is almost constant at $\tau$(\hi)=0.8 over an
angular extent of at least 0.7 arcsec. Fig.~\ref{3c161taumap131} shows
two prominent peaks at a velocity of 11.5 km s$^{-1}$ at angular
displacements of 1.8 and 1.4 arcsec, with excesses in $\tau$(\hi) of
0.5 and 0.6 and angular sizes of 0.5 and 0.2 arcsec (400 and 160 AU),
respectively. The corresponding spectral feature has a FWHM of $3.2\pm0.3$ km s$^{-1}$.

The \hi\/ opacity at 20.2 km s$^{-1}$ (Fig.~\ref{3c161taumap110}) is
an order of magnitude weaker than at lower velocities.  This spectral
feature has a rather uneven shape with a FWHM of $2.5\pm0.5$ km s$^{-1}$. The slice
shows at least one \hi\/ component at an offset of 1.7 arcsec with
fluctuations in $\tau$(\hi) of $\approx0.1$ and an angular size of 0.5
arcsec (900 AU at a distance of 1.8 kpc).  The highest velocity and
hence most distant \hi, at 34.2 km s$^{-1}$
(Fig.~\ref{3c161taumap76}), has a well-defined spectral profile of
FWHM $3.0\pm0.2$ km s$^{-1}$. The line does not show evidence of any
significant variation in the \hi\/ opacity over an angular scale of
$\approx$ 1 arcsec (3500 AU at a distance of 3.5 kpc).

\section{Summary and Conclusions} 
\label{concl}
We have used the MERLIN array to image Galactic \hi\/ absorption, with
a resolution of $\sim$ 0.1~arcsec, in the direction of three
background sources, 3C~111, 3C~123 and 3C~161. We resolve the \hi\/
distribution towards the latter two objects with angular sizes $\geq$
1~arcsec.
In the case of  3C~161,
prominent spatial variations at velocities close to zero (in the range
4.5--11.5 km s$^{-1}$) have
been observed.  In addition, at  20.2 km s$^{-1}$ the
opacity of the \hi\/ is an order of magnitude lower
($\approx0.1-0.2$);  at this velocity significant fluctuations in the
\hi\/ opacity across the face of 3C161 are observed. The inferred
distance of this \hi\/ feature is 1.8 kpc. At the highest velocity in
this direction (34.2 km s$^{-1}$ at an inferred distance of 3.5 kpc) no
significant changes in the weak \hi\/ features with
$\tau$(\hi)$\approx0.1$ are observed.  In the direction of 3C123, there
are suggestions of opacity variations in the low-velocity \hi\/, but
with a much lower statistical significance.  

 As pointed out by \citet{gl86}, the expectation would be that the
\hi\/ features at higher velocities would subtend smaller solid angles
and thus cover a smaller fraction of the continuum background source;
however the existence of a cloud size spectrum and random cloud
velocities could mask the expected behaviour in this small sample. In
addition, the higher-velocity \hi\/ lines have smaller opacities (0.1
to 0.2) compared to the lower velocity lines ($\tau \sim$ 1). This
decrease in mean opacity reduces the sensitivity of these studies with
respect to detecting significant spatial variations in the higher
velocity lines.

On a spatial scale of 400 to 500 AU, prominent
changes in opacity at the level of $\delta\tau\sim 1$ in
the direction of  3C~161 are striking. \citet{brog05} had
previously determined $\delta\tau\sim 0.5$ with a linear resolution of
$\sim$ 25 AU in the direction of 3C~138.
The peak change in $\delta\tau$ of about 1 seen in the MERLIN data
is consistent with the peak fluctuation of about 0.7 on scales of 400--500 AU 
predicted by \citet{desh00}.
Previous studies of small scale \hi\/
have traditionally calculated the implied densities based on the
assumption that the line of sight dimension is comparable to the
observed transverse dimension.  \citet{heiles07} has discussed this
assumption in detail; the expectation is that the objects are far from
isotropic. 
The 400--500 AU structures detected in these observations
would imply volume densities of $< 5 \times 10^{3}$ cm$^{-3}$ with column
densities a few times $10^{20}$ cm$^{-2}$ if the excitation
temperature is 50~K \citep{brog07} and the objects have a path length along the line of sight that is more than ten times the transverse dimension.

\section{Acknowledgements} 
We thank Kristina Barkume, who began the research on the 3C~161 data as
a part of a senior thesis at Reed College in 2002--2003, and Michael
Faison, who participated in the initial phases of this project. The
National Radio Astronomy Observatory is a facility of the National
Science Foundation operated under a cooperative agreement by
Associated Universities, Inc. MERLIN is operated by the University of
Manchester on behalf of the STFC (formerly PPARC).

\clearpage \begin{table*}
\caption{Source and Observing Parameters. The duration is the total
length of each observation, about 2/3 of which was spent on the main
target. The RMS noise $\sigma_{\rm rms}$ given for the line
observations is for a single spectral channel.  \label{obspar}}
\begin{tabular}{|l|cccc|} 
\hline
{Source}& & 3C~111 & 3C~123 &3C~161\\
Galactic  Coordinates&    (\emph{l\degr, b\degr})&161.68, --8.82  &	170.58, --11.66		& 215.44, --8.07 \\
\multicolumn{2}{l}{\bf Broad-band Continuum} &&&\\
Date & YYYMMDD &   20030202 & 20030204 & 20011030\\
Duration & (hr) & 17.5& 16 & 10\\
Frequency & (MHz) & 1408  & 1408 & 4994\\
Phase Reference Source & & B0415+398 &B0430+289 &B0605-085\\
Target Peak &(Jy bm$^{-1}$)&1.285&0.749&0.539\\
RMS noise & (mJy bm$^{-1}$)&2.0&2.6&0.6\\
\multicolumn{2}{l}{\bf Narrow-band/Line}&&&\\
Date & YYYMMDD &20030508 & 20030501&20020304\\
Duration & (hr)&16&16&9\\
 $V_{\rm LSR}$&(km s$^{-1}$) &5.0 & 3.0 &17.5\\
Phase Reference Source & & B0415+398 &B0436+306 &B0605-085\\
Phase Ref. Flux Density & (Jy bm$^{-1}$) & 0.41  & 0.55 & 2.1\\
Synthesized Beam & (arcsec$\times$arcsec) &0.19 $\times$ 0.15& 0.23 $\times$ 0.14 & 0.46 $\times$ 0.13 \\
Position Angle &(deg)& 45&28&21\\
Continuum Peak &(Jy bm$^{-1}$) & 1.27& 0.75 & 3.40\\
Continuum $\sigma_{\rm rms}$ &(mJy bm$^{-1}$)&2&4&4\\
Line $\sigma_{\rm rms}$ &(mJy bm$^{-1}$) & 8.5 & 11 & 18\\
\hline
\end{tabular}
\end{table*}

\clearpage

 \clearpage

 \clearpage

 \clearpage

 \clearpage

 \clearpage

 \clearpage

 \clearpage

 \clearpage

 \clearpage

 \clearpage

 \clearpage

 \clearpage

\appendix
\section{Continuum observations of 3C161}
\label{appendix}

We describe the radio continuum images of 3C161 obtained as part of
this work since they are not published fully elsewhere. Although a
powerful radio source, 3C161 lies at low Galactic latitude ($b=-8.07$\degr)
and has therefore escaped much attention since it has often been excluded
from samples under study. \citet{gl86} showed that this
source possesses a classical double radio structure with an angular
size of 62 arcseconds, albeit with a large brightness ratio in excess
of 60 between the Northern and Southern hotspots. The \hi\/ absorption
reported in this paper is associated with the brighter Northern
hotspot.

\begin{figure*} \centering \includegraphics[angle=0,width=18cm]{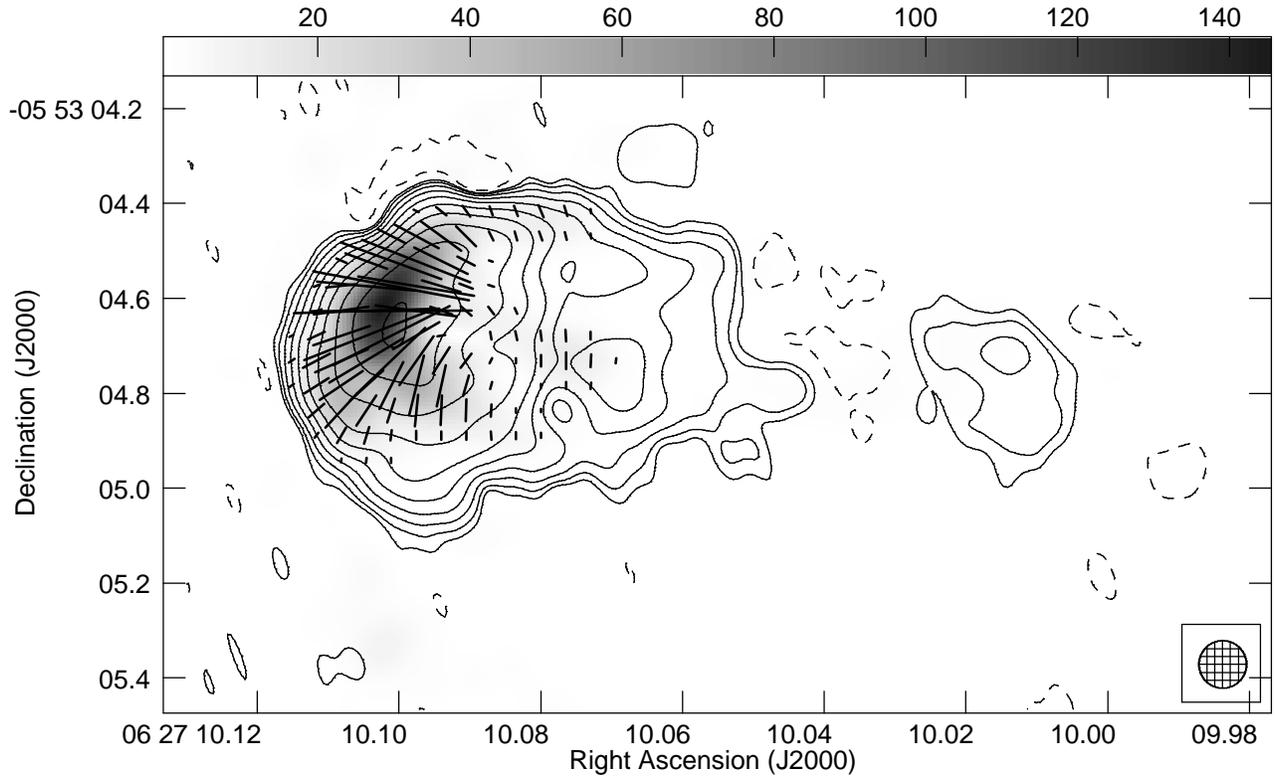} \caption{MERLIN continuum image
of 3C~161 observed at 4994 MHz using 15-MHz bandwidth.  The contour
levels are (--3, 3, 6, 12, 24 ...) $\times$ the rms noise level of
0.35 mJy beam$^{-1}$.  The circular beam of 100 mas FWHM is indicated
in the lower right corner. The greyscale shows the polarized intensity
in mJy beam$^{-1}$. The vectors represent the direction of the {\bf E}
field and 10 mas length corresponds to 4 mJy beam$^{-1}$ polarized
flux density.  } \label{3C161_MER_5.PS} \end{figure*}

 A MERLIN high-resolution 5-GHz image is shown in
Fig.~\ref{3C161_MER_5.PS}. The observations are summarised in
Table~\ref{obspar} and the data reduction is outlined in
Section~\ref{obs}.  The point source calibrator, B0552+398, was found
to have a flux density of 5.74 Jy by comparison with 3C286. 3C~161 and
the nearby (5.5\degr) phase-reference source B0605-085 were observed
alternately for 5.5 and 2 min, respectively, over a total of 9.5
hr. We also used B0605-085 to correct for polarization leakage and we
calibrated the polarization angle (P.A.) with respect to 3C286 which
has an assumed P.A. of +33\degr\/ (in the map plane).

Figure~\ref{3C161_MER_5.PS} shows a conventional hotspot structure
(peak 539 mJy beam$^{-1}$). The total intensity map
noise around the hotspot is increased about three-fold with respect to
more distant regions of the map (and the polarized intensity images)
due to sidelobes of this bright source.

\begin{figure*} \centering \includegraphics[angle=0,width=18cm]{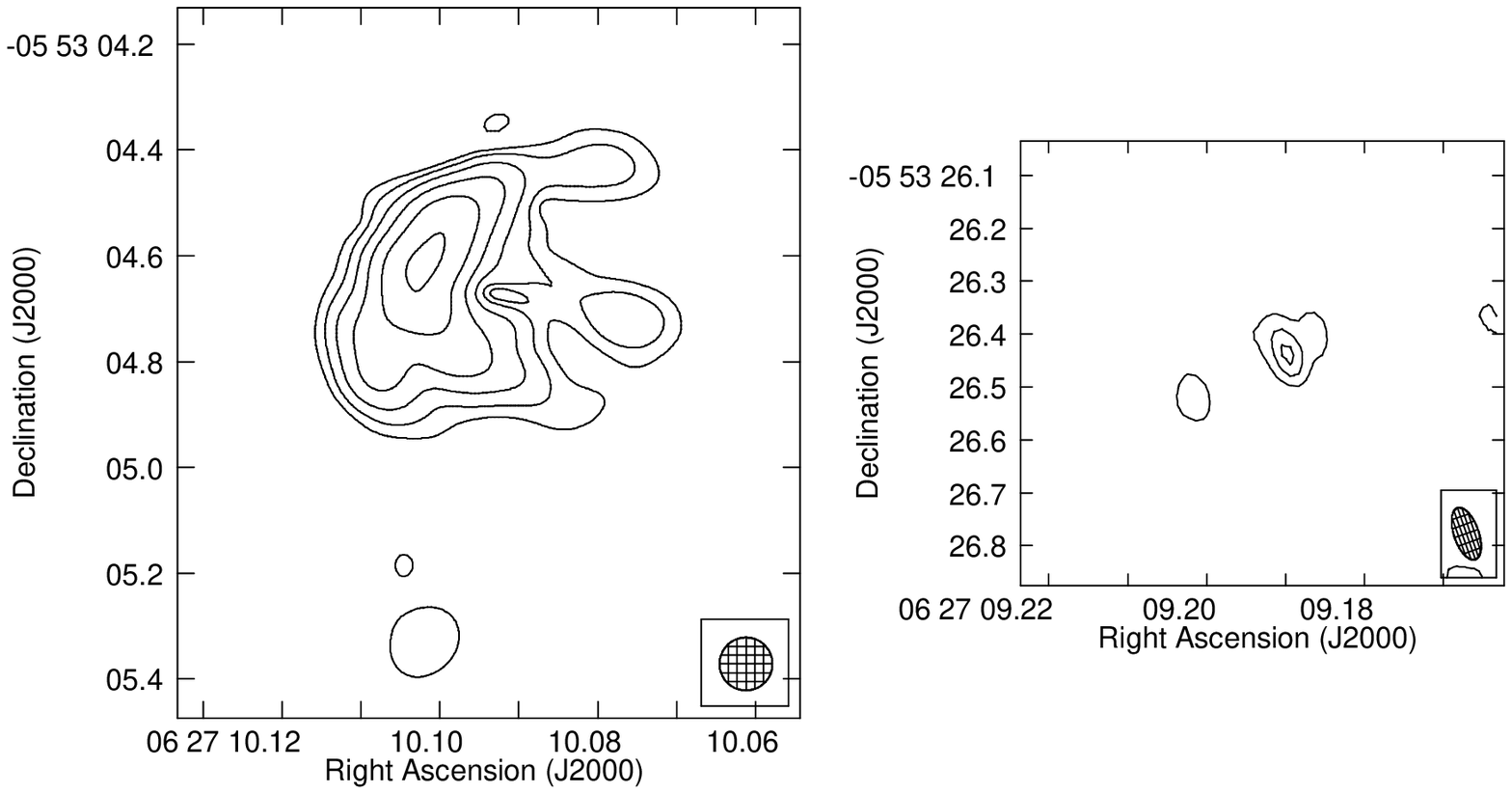} \caption{MERLIN continuum images of 3C~161 observed at 4994 MHz using 15-MHz bandwidth. {\em left}  Polarized intensity.
The contour levels are (-1,1, 2, 4 ... 32) $\times$ 4 mJy beam$^{-1}$. 
The circular  beam of 100 mas FWHM is indicated in the 
lower right corner. {\em right}  The core candidate in total intensity.
 The contour levels are (--3, 3, 6, 9)  $\times$ the local rms noise
 level of 0.12  mJy beam$^{-1}$. The   beam is $103$ mas $\times 47$ mas at a
 position angle of 19\degr, indicated in the lower right corner.
} \label{3C161_MER_POLI+CORE.PS} \end{figure*}

The polarized intensity image, Fig.~\ref{3C161_MER_POLI+CORE.PS} ({\em
left}), shows that the brightest polarized emission lies around the
outer edge of the hotspot region.  The maximum polarized intensity
detected is 145 mJy beam$^{-1}$.  The plotted vectors indicate the
directions of the {\em in situ} {\bf E} field, which have been
corrected for foreground rotation using the rotation measure of 112
rad m$^2$ given by \citet{Vallee88}. These vectors indicate a
circumferential magnetic field, as would be expected around the outer
edge of the working surface of a lobe.  The trajectory of the incoming
beam or jet can be traced in the polarized intensity image,
Fig.~\ref{3C161_MER_POLI+CORE.PS} ({\em left}). The jet enters the
hotspot region from the West where the {\bf E} vectors in
Fig.~\ref{3C161_MER_5.PS} are parallel to each other and perpendicular
to the jet direction (implying a magnetic field aligned with the
jet). The fractional polarization is highest here and at the leading
edge of the lobe, at $\sim30$ per cent.  The polarized intensity falls to nearly zero just
as the jet enters the hotspot in a region where the polarized position
angle is rotating rapidly, presumably due to shocks.

In addition to the bright emission associated with the Northern
hotspot, we detect a  compact component of $1.15\pm0.12$ mJy beam$^{-1}$
($10\sigma$ at the local rms noise level) at a distance of 25.5
arcsec SW of the hotspot, shown in
Fig.~\ref{3C161_MER_POLI+CORE.PS} ({\em right}). Its position is J~2000 right
ascension $06^{\rm h} 27^{\rm m}
09\fs189$, declination $-05\degr 53\arcmin 26\farcs43$ (uncertainty 25
mas).
This is a candidate
for the core of 3C~161; however, we note that it is not seen in any
other available data. MERLIN did not detect any  emission from
the much weaker Southern hotspot.

\begin{figure*} \centering \includegraphics[angle=0,width=18cm]{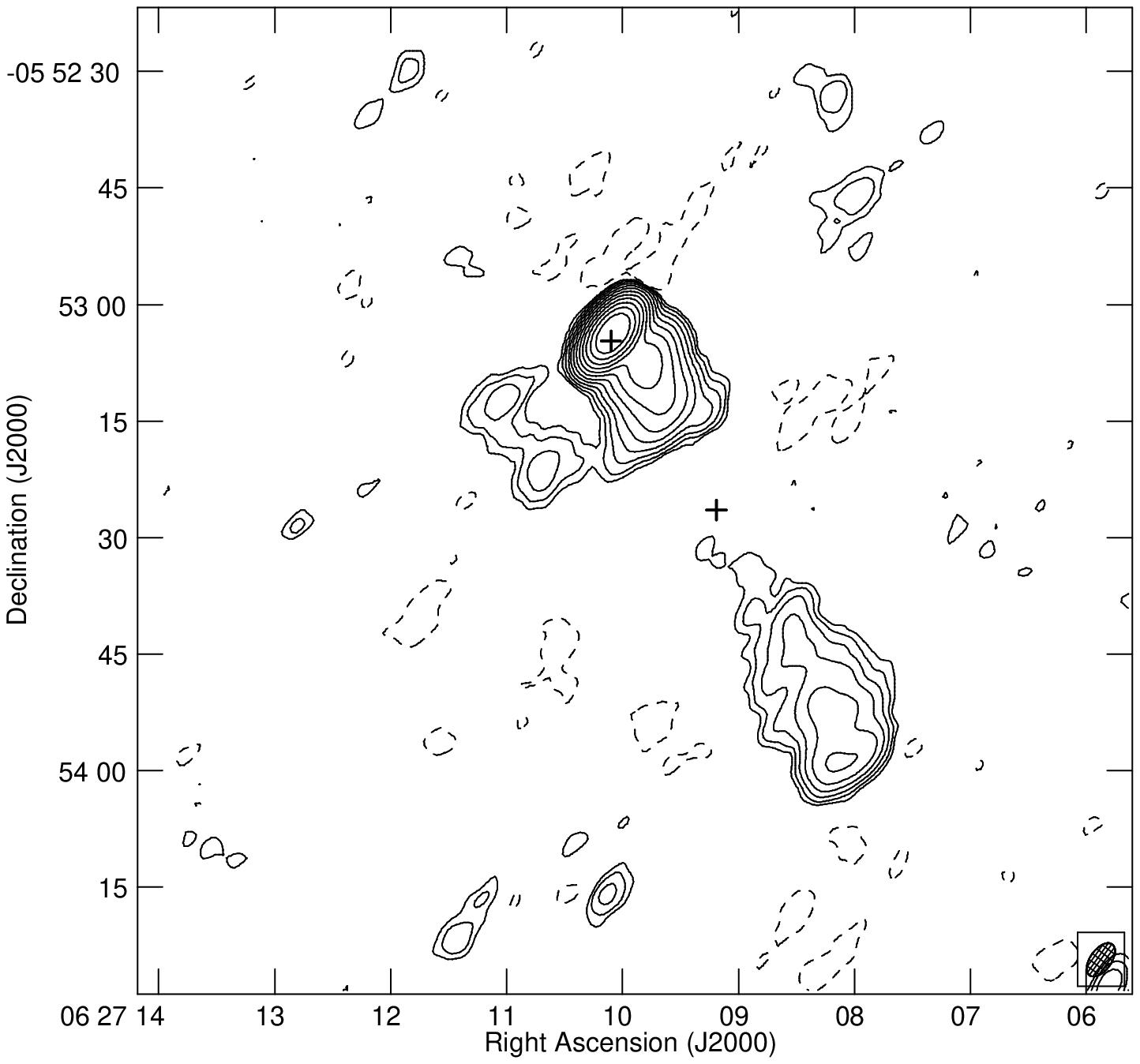} \caption{VLA continuum image
of 3C~161 observed at 4860 MHz using 100-MHz bandwidth.  The contour
levels are (--3, 3, 6, 12, 24 ...) $\times$
the rms noise level of 0.33 mJy beam$^{-1}$. The  beam is 4\farcs94
$\times$ 2\farcs71 at a position angle of --36\degr, indicated in the lower right corner. The crosses mark the positions of the MERLIN hotspot peak and the core candidate.
} \label{3C161_VLA_C.PS} \end{figure*}

Figure~\ref{3C161_VLA_C.PS} shows an image of 3C~161 at 5 GHz which we
made from VLA archive data (proposal code AP370, observed on 19980915
in the B-array configuration).  The source was observed as a
calibrator and, to our knowledge, no maps at this frequency and
resolution range have previously been published. The positions of the
MERLIN hotspot and core candidate are marked. The VLA did not detect
any core emission but the 3$\sigma$ noise level of this image is close
to the peak emission of the core candidate measured by MERLIN, so it
is feasible that it was just below the detection threshold of these VLA data.

\begin{figure*} \centering \includegraphics[angle=0,width=18cm]{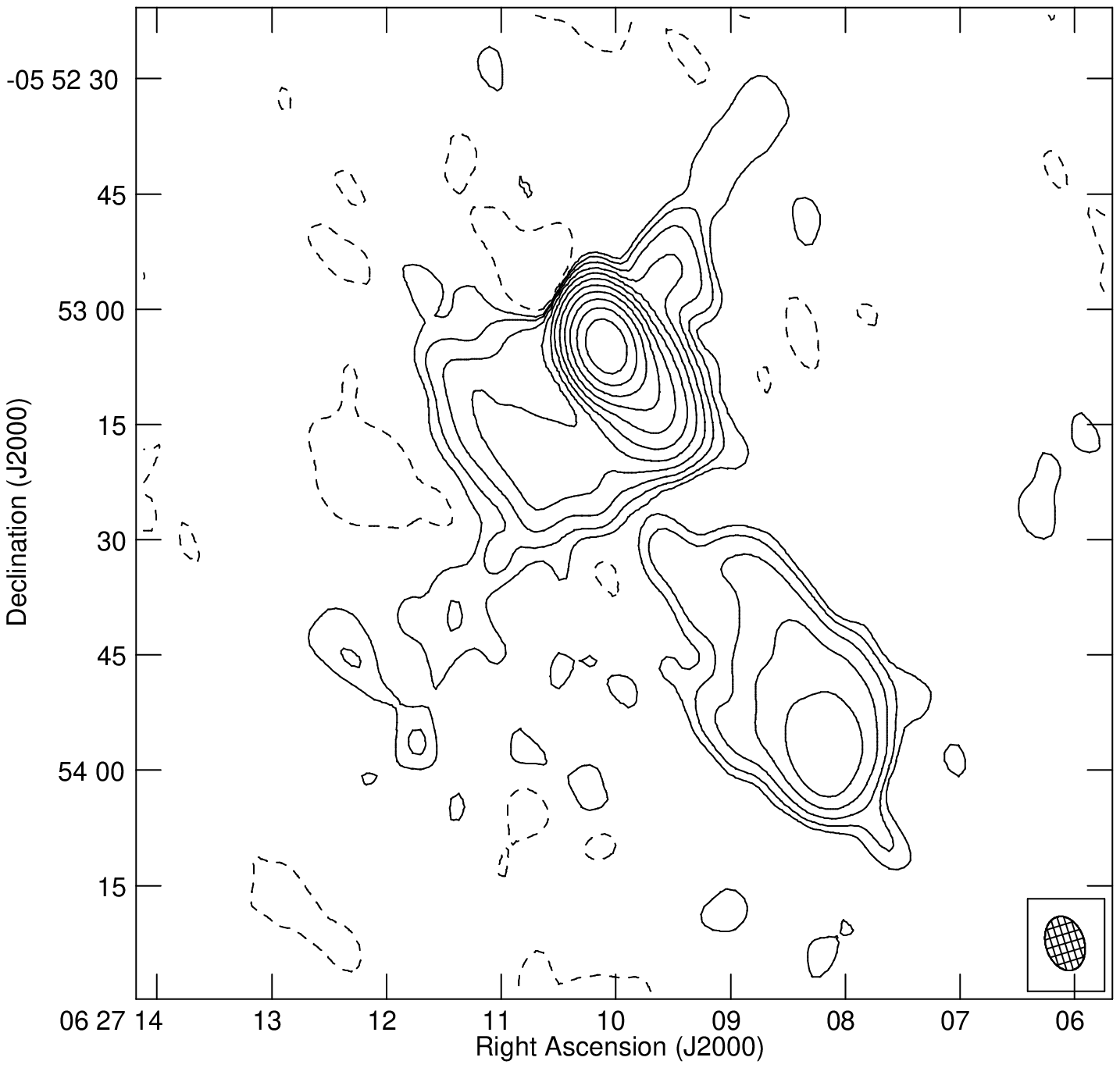} \caption{VLA continuum image
of 3C~161 observed at 1420 MHz using 0.15-MHz bandwidth.  The contour
levels are (--3, 3, 6, 12, 24 ... ) $\times$
the rms noise level of 2 mJy beam$^{-1}$. The beam is 7\farcs19
$\times$ 5\farcs01 at a position angle of 17\degr, indicated in the lower right corner.
} \label{3C161_VLA_L.PS} \end{figure*}

 Figure~\ref{3C161_VLA_L.PS} shows the 1.4-GHz continuum counterpart
to the \hi\/ data published by \citet{gl86}. The data have been
re-processed and the image shown here is naturally-weighted. This map is
thus somewhat deeper and more sensitive to extended low-surface-brightness
emission than the original, figure 2a in \citet{gl86}. The Northern
component of this FRII double source is a little unusual since it has
a tail which arcs around through $\sim180$ degrees to the SE. This
is in the same sense as the jet entry direction into the
hotspot. There is also low surface brightness emission SW of the double structure which may be associated with the
source.

A linear feature, seen in Fig.~\ref{3C161_VLA_L.PS}
extending from NW of the Northern hotspot, is difficult to
interpret.  It could be associated with a faint optical object seen in
the Digitised Sky Survey\footnote{\tt http://archive.stsci.edu/dss/index.html} which also lies just  NW
of the 3C161 hotspot, but this cannot be related to 3C161.  Indeed, there are
no obvious optical counterparts which can be related to any part of the observed radio structure of 3C161, and thus it remains unidentified.

 \clearpage

 \clearpage

 \clearpage


\end{document}